\documentclass[aps,prd,nofootinbib,superscriptaddress]{revtex4-2} 
\usepackage{amsmath,amssymb,graphicx,hyperref,xcolor,bm}
\usepackage{booktabs}
\usepackage{multirow}

\begin{document}

\title{
A Unified Interpretation of Supernova, GRB, and QSO Time Dilation Signals\\
in a Generalized Cosmological Time Framework
}

\author{Seokcheon Lee}
\affiliation{Department of Physics, Institute of Basic Science, Sungkyunkwan University, Suwon 16419, Korea}

\date{\today}


\begin{abstract}
Cosmological time dilation (CTD) serves as a fundamental probe of cosmic expansion, historically verified through the characteristic $(1+z)$ broadening of Type Ia supernova (SNe Ia) light curves. However, significant tensions arise when extending this test to other astrophysical regimes. While discrete, event-based transients such as Gamma-Ray Bursts (GRBs) exhibit large scatter in inferred time-dilation signatures, analyses of stochastic variability in persistent sources—specifically Quasars (QSOs)—frequently yield null results. I demonstrate that these discrepancies stem from a previously overlooked distinction between discrete geometric clocks and continuous thermal emission, presenting a resolution within the framework of Generalized Cosmological Time (GCT). The central premise relies on strictly distinguishing global coordinate time, characterized by a generalized lapse function, from the local proper time measured within gravitationally bound systems. I propose that the progenitors of transients—specifically SNe~Ia and GRB central engines—are effectively shielded from background time evolution due to strong gravitational binding and environmental decoupling. 
Consequently, they act as standard clocks tracing pure geometric path dilation, obeying $\tau_{\rm obs} \propto (1+z)^{1-b/4}$. Conversely, the lack of dilation in QSOs is derived as a consequence of observing persistent thermal accretion disks at fixed wavelengths, introducing an intrinsic selection effect ($\tau_{\rm intr} \propto (1+z)^{-2}$) that masks the cosmological signal. This framework reconciles the diverse behaviors of transient and persistent sources without modifying local physical laws.
\end{abstract}

\maketitle

\tableofcontents

\section{Introduction}
\label{sec:intro}

Cosmological Time Dilation (CTD) stands as one of the most direct and fundamental observational consequences of the expanding universe. In the standard Friedmann-Lema\^{i}tre-Robertson-Walker (FLRW) cosmology, a physical timescale $t_{\rm em}$ at emission is observed as $t_{\rm obs}=t_{\rm em}(1+z)$, independent of the matter content or the cosmic distance ladder. Consequently, CTD measurements offer a particularly clean probe of cosmic kinematics, complementing distance-based observables such as luminosity distances or baryon acoustic oscillations (BAOs)~\cite{Lee:2023ucu,Lee:2023rqv,Lee:2021ona,Lee:2024kxa}.

Despite the theoretical simplicity of CTD, observational tests across different classes of astrophysical sources yield markedly different results, presenting a significant puzzle. Event-based transients, such as Type~Ia supernovae (SNe~Ia), exhibit a clear $(1+z)$ stretching of their light-curve durations and spectral evolution, consistent with relativistic expansion~\cite{Leibundgut:1996qm,SupernovaSearchTeam:1997gem,Foley:2005qu,Blondin:2007ua,Blondin:2008mz,DES:2024vgg}. By contrast, Gamma-Ray Bursts (GRBs) show weaker and highly scattered evidence for time dilation, with observed durations strongly influenced by intrinsic engine variability and energy-dependent definitions of $T_{90}$~\cite{Kouveliotou:1993yx,Qin:2012ht,Shahmoradi:2012mp,Bargiacchi:2024srw}. Even more striking are studies of the stochastic variability in persistent sources, specifically Quasars (QSOs), which frequently report a null result implying little or no measurable CTD signal over a wide redshift range~\cite{Hawkins:2001be,Hawkins:2010xg,OGLE:2009crb,Caplar:2017apj}. Although recent hierarchical modeling has hinted at time dilation in QSOs~\cite{Lewis:2023jab}, the discrepancy between the standard $(1+z)$ prediction and raw variability data remains a significant interpretational challenge.

In this work, I demonstrate that these disparate observational behaviors can be unified within the framework of Generalized Cosmological Time (GCT)~\cite{Lee:2020zts,Lee:2023bjz}. The framework, originally introduced as the minimally extended Varying Speed of Light (meVSL) model, is naturally interpreted here not as a modification of local relativistic physics, but as a generalized choice of the cosmological time gauge. Recent studies have clarified that the apparent variation of dimensional constants in the Robertson-Walker (RW) spacetime reflects a gauge freedom associated with the choice of the lapse function $N(t)$ in the metric~\cite{Ryder:2009GR,Lee:2024zcu,Lee:2025osx}. The theoretical foundation of this analysis rests on the rigorous distinction between the global coordinate time $t$ and the proper time $\tau$ measured by a comoving observer. In General Relativity (GR), these two timescales are related via the lapse function, $d\tau = N(t)dt$. While the standard cosmological model implicitly adopts the gauge $N(t)=1$, identifying coordinate time with proper time, there is no a priori reason to enforce this choice globally~\cite{Arnowitt:1962hi,Misner:1973prb,Weinberg:2008zzc}. The GCT framework explores the physical consequences of a generalized temporal gauge $N(t) \propto a^{b/4}$, where the coordinate clock rate governing the expansion differs from the local proper clock rate governing atomic transitions~\cite{Lee:2022heb}. Within this interpretation, local Lorentz invariance and all local physical laws—including special relativity, quantum mechanics, and local field theory—remain strictly preserved when measured by proper time.

The central hypothesis of this paper is the principle of environmental decoupling. I propose that gravitationally bound systems—specifically the white dwarf progenitors of SNe~Ia and the central engines of GRBs—are dynamically decoupled, or ``shielded,'' from the background cosmological time evolution. Just as the spatial expansion scale factor $a(t)$ does not stretch the orbits of atoms or solar systems, the generalized cosmological time flow implies that the lapse function gradient does not penetrate the deep gravitational potential wells of compact sources where local dynamics are governed by the Schwarzschild-like metric. Consequently, SNe~Ia and GRBs act as pure background probes where intrinsic proper timescales are fixed, and the observed dilation arises purely from the photon path through the expanding metric. By incorporating the lapse function, the observed coordinate duration obeys a generalized scaling $t_{\rm obs} \propto (1+z)^{1-b/4}$. In this context, the parameter $b$ is not merely a gauge artifact but serves as a measurable physical parameter that quantifies the accumulated difference between the background cosmological clock rate and the local atomic clock rate.

In sharp contrast, the apparent lack of dilation in QSOs is identified as a consequence of observing a continuous source at a fixed observed wavelength. I demonstrate that for thermally stratified accretion disks, observing at a fixed bandpass selectively probes inner, faster-dynamical regions at higher redshifts. Since the photon wavelength scaling $\lambda_{\rm obs} = \lambda_{\rm rest}(1+z)$ remains valid in the GCT framework, a fixed $\lambda_{\rm obs}$ corresponds to a rest-frame wavelength $\lambda_{\rm rest} \propto (1+z)^{-1}$. Assuming standard accretion disk physics, the characteristic variability timescale—measured in the source's proper time—follows the scaling $\tau_{\rm rest} \propto \lambda_{\rm rest}^2$ (derived from the Keplerian relation $\tau \propto R^{3/2}$ and the thermal profile $T \propto R^{-3/4}$). This introduces an intrinsic proper-time shortening effect $\tau_{\rm rest} \propto (1+z)^{-2}$. This strong intrinsic evolution effectively masks the geometric time dilation signal ($t_{\rm obs} \propto \tau_{\rm rest}(1+z)^{1-b/4}$), creating the illusion of a non-dilated source.

This geometric reinterpretation provides a consistent resolution to several observational puzzles beyond CTD. In particular, the mild redshift evolution inferred in the local Hubble parameter $H_0(z)$ from SNe~Ia data can be understood as a consequence of a gauge-dependent normalization of cosmological time, rather than as evidence for modified expansion dynamics~\cite{Lee:2025arX}. Ref.~\cite{Lee:2025vha} further demonstrated that generalized CTD simultaneously shortens the sound horizon at the drag epoch and modifies transient time dilation, providing a self-consistent mechanism to alleviate the Hubble tension with a single parameter $b$, which quantifies the relative normalization between local proper time and cosmological time hypersurfaces. In this respect, the present approach is conceptually distinct from look-back-time-based discussions of the $H_0$ tension, which operate at the level of integrated cosmic ages. Rather than modifying the look-back time itself, the GCT framework focuses on the operational definition of physical clocks and their coupling to the background time slicing. From this perspective, the generalized clock behavior discussed here provides a natural interpretive underpinning for the effective redefinitions of
look-back time advocated in previous studies~\cite{Capozziello:2023ewq,Vagnozzi:2023nrq,Capozziello:2024stm,Hu:2025fsz}.

By systematically applying these principles, this paper demonstrates that SNe~Ia, GRBs, and QSOs are not contradictory; rather, they represent different classes of clocks—shielded versus bandpass-selected—interacting with the same expanding background metric. The remainder of this paper is organized as follows. Section~\ref{sec:theoretical_framework} establishes the theoretical foundations, interpreting the GCT framework as a generalized temporal gauge and deriving the environmental shielding mechanism that dynamically decouples gravitationally bound systems from the background time evolution. Section~\ref{sec:duration_scaling} then applies this framework to observational data, validating the generalized dilation law with SNe~Ia and GRBs while analytically deriving the selection effect that masks the signal in QSOs. Finally, Section~\ref{sec:discussion} discusses the broader implications for cosmological tensions, and Section~\ref{sec:conclusion} summarizes the findings.

\section{Theoretical Framework: Generalized Cosmological Time and Environmental Shielding}
\label{sec:theoretical_framework}

This section establishes the theoretical foundation of the analysis, rigorously interpreting the construction previously referred to as the meVSL model within the GCT framework. The distinction is fundamental: this framework proposes neither a modification of local relativistic physics nor a violation of Lorentz covariance. Instead, it explores the physical consequences of a generalized temporal gauge choice within the RW metric~\cite{Wald:1984rg,Gourgoulhon:2007ue}. By strictly distinguishing between the global coordinate time characterizing the expansion and the local proper time governing bound systems, I derive the mechanism of \textit{environmental shielding}, which acts as the physical basis for the disparate behaviors of astrophysical clocks.

\subsection{Gauge Freedom in the Robertson-Walker Metric}

Within GR, full diffeomorphism invariance permits the lapse function $N(t)$ of the RW metric to be fixed by an arbitrary coordinate choice~\cite{Arnowitt:1962hi,Misner:1973prb,Weinberg:2008zzc}. The invariant line element is given by
\begin{equation}
    ds^2 = -N^2(t) c_0^2 dt^2 + a^2(t) d\vec{x}^2 = -c_0^2 d \tau^2 \,.
\end{equation}
The standard cosmological model implicitly adopts the specific temporal gauge $N(t)=1$, identifying coordinate time $t$ directly with the proper time $\tau$ of comoving observers. However, GR provides no first-principles argument that this synchronous gauge uniquely corresponds to the physical time realized across the expanding universe~\cite{Ryder:2009GR,Ellis:2012GR}. Importantly, the generalized scaling introduced in the GCT framework does not modify the background expansion dynamics governed by the Friedmann equations. Matter sources affect the cosmic expansion history in the standard manner, and no additional dynamical degrees of freedom are introduced. The parameter $b$ characterizes a temporal gauge choice within the RW metric, not a modification of the energy-momentum content or gravitational field equations.

In particular, background observables that depend on the combination $c/H$ remain strictly invariant under this generalized temporal gauge. For example, the comoving distance (equivalently, the conformal look-back integral) is given by~\cite{Lee:2020zts,Lee:2024zcu,Lee:2025osx,Lee:2025arX}
\begin{equation}
\chi(z) \;=\; \int_0^z \frac{c(z')}{H(z')} \, dz'
\;=\; \int_0^z \frac{c_0}{H^{(\rm{SM)}}(z')} \, dz' \, .
\end{equation}
Here the gauge-dependent scaling of $c(z)$ is exactly compensated by the
corresponding scaling of $H(z)$ defined with respect to the same generalized time
coordinate. This equality makes explicit that the background geometry and
distance--redshift relations are mathematically equivalent to those of the standard FLRW cosmological model, and that the GCT framework represents a temporal gauge choice rather
than a modification sourced by matter.Consistently, the look-back time expressed in proper time is identical to that obtained in the standard FLRW formulation
\begin{align}
\int^{\tau_0}_{\tau_{\rm{em}}} d \tau = \int^{t_0}_{t_{\rm{em}}} N(t) dt = \int^{1}_{a_{\rm{em}}} \frac{N da}{aH} = \int^{1}_{a_{\rm{em}}} \frac{N da}{a N H^{(\rm{SM)}}} = \int_{0}^{z} \frac{dz'}{(1+z') H^{(\rm{SM})}(z')} \label{tlb} \, 
\end{align}
where we have used $d\tau = N(t)\,dt$ and $H \equiv \frac{1}{a}\frac{da}{dt} = N H^{(\rm SM)}$ defined with respect to the generalized time coordinate.

In this analysis, I adopt the GCT framework as a controlled generalization where the lapse function scales as $N(t) \equiv a^{b/4}$. Here, the parameter $b$ quantifies the deviation of the cosmological clock rate from the standard $N=1$ slicing. The apparent redshift dependence of dimensional quantities often associated with this metric—such as a varying speed of light $c(z) = c_0 a^{b/4}$ or Planck constant $\hbar(z) = \hbar_0 a^{-b/4}$—is strictly a coordinate artifact arising from the gauge choice. As summarized in Table~\ref{tab:gct_scalings}, these scalings are not arbitrary but are uniquely fixed by the simultaneous requirement of RW symmetry~\cite{Lee:2022heb} and the preservation of local physics~\cite{Lee:2020zts,Lee:2024zcu,Lee:2025osx,Lee:2025arX}. For instance, while coordinate values evolve, the locally measured relations such as $E=mc^2$ and $E=\hbar\nu$ remain strictly invariant within the local frame.

\begin{table}[htbp]
\centering
\resizebox{\textwidth}{!}{%
\begin{tabular}{c c c c}
\toprule
\textbf{Sector} 
& \textbf{Representative quantities} 
& \textbf{GCT scaling with } $a$ 
& \textbf{Preserved local relation (Shielded)} \\
\midrule
Relativistic kinematics
& mass, light speed 
& $m(a)=m_0 a^{-b/2}, \quad c(a)=c_0 a^{b/4}$ 
& $E=mc^2$ \\

Gravitational coupling
& Newton constant 
& $G(a)=G_0 a^{b}$ 
& $c^4/G=\text{const}$ \\

\multirow{2}{*}{Electromagnetic sector}
& charge, frequency, wavelength 
& $e(a)=e_0 a^{-b/4},\; \nu(a)=\nu_0 a^{-1+b/4},\; \lambda(a)=\lambda_0 a$ 
& $\lambda\nu=c$ \\
& permeability, permittivity
& $\mu(a) = \mu_0 a^{-b/4}, \quad \epsilon(a) = \epsilon_0 a^{-b/4}$
& $c^2 = 1/(\mu\epsilon)$ \\

Quantum relations
& Planck constant 
& $\hbar(a)=\hbar_0 a^{-b/4}$ 
& $E=\hbar\nu$ \\

Thermodynamics
& temperature 
& $T(a)=T_0 a^{-1}$ 
& $E=k_{\rm B}T$ \\
\bottomrule
\end{tabular}
}
\caption{
Correlated cosmological scalings of representative dimensional quantities in the GCT framework. The apparent redshift dependence arises from a generalized temporal gauge choice in the RW metric ($N(t) = a^{b/4}$). While the coordinate values scale with $a$, all physical relations measured in the local proper frame of a shielded system remain invariant. Note that the relation $c^2 = 1/(\mu\epsilon)$ is preserved as $c^2 \propto a^{b/2}$ and $(\mu\epsilon)^{-1} \propto (a^{-b/4}a^{-b/4})^{-1} = a^{b/2}$.
}
\label{tab:gct_scalings}
\end{table}

\subsection{The Mechanism of Environmental Shielding}
\label{subsec:shielding}

A fundamental question arises regarding the consistency of this framework: if the background spacetime is characterized by a time-varying lapse function, why do local atomic clocks or nuclear decay rates at high redshift not exhibit this shift? The resolution lies in the principle of \textit{environmental shielding}, which is the temporal analogue of the well-established spatial decoupling of bound systems.

It is a standard result in cosmology that the spatial expansion of the universe, described by the scale factor $a(t)$, does not extend to gravitationally bound systems. Atoms, solar systems, and galaxies do not expand with the Hubble flow because their local dynamics are governed by electromagnetic or gravitational binding forces rather than the global FLRW metric. By symmetry, I propose that this decoupling extends to the temporal domain. Just as the spatial metric $g_{ij}$ decouples from the cosmic expansion $a(t)$ within a virialized system, the temporal component $g_{00}$ decouples from the cosmic lapse function $N(t)$.

This concept is grounded in the Strong Equivalence Principle (SEP) and realized via the McVittie solution, which describes a local mass embedded in an expanding universe~\cite{McVittie:1933zz}. Inside a virialized region where the local Newtonian potential $\Phi(r)$ dominates ($r \ll H^{-1}$), the effective metric segregates from the global expansion. Of particular significance is that the effective lapse function in this bound region is static to first order, $N_{\text{eff}} \approx 1 - \Phi/c_0^2$. This implies that the bound system effectively maintains a static gauge where the coordinate time flow synchronizes with the local proper time, regardless of the global $b$ parameter ($b_{\text{local}} \approx 0$). Consequently, the local physics is shielded against the background evolution, ensuring that intrinsic timescales $\tau_{\rm rest}$—such as the decay time of Nickel-56 in SNe or the accretion timescale in GRBs—are constant across cosmic history
\begin{equation}
\tau_{\rm rest}(z) \approx \tau_{\rm rest}(0) = \text{const}.
\end{equation}

\subsection{Observables: Path Dilation vs. Intrinsic Selection}
\label{subsec:observables}

The distinction between the shielded local frame and the expanding background frame leads to specific observational consequences for CTD. While the source itself is shielded, the photons emitted must traverse the expanding background metric to reach the observer. To derive the observable signature rigorously, consider the propagation of light along a null geodesic ($ds^2=0$). From the line element, the comoving distance traversed by a photon emitted at coordinate time $t_{\rm em}$ and observed at $t_{\rm obs}$ is given by
\begin{equation}
    \Delta x = \int_{t_{\rm em}}^{t_{\rm obs}} \frac{c_0 N(t)}{a(t)} dt = \text{constant}.
\end{equation}
Since the comoving coordinate distance to the source is fixed, differentiating this integral with respect to the emission and observation times yields the relation between the time intervals
\begin{equation}
    \frac{c_0 N(t_{\rm obs})}{a(t_{\rm obs})} dt_{\rm obs} - \frac{c_0 N(t_{\rm em})}{a(t_{\rm em})} dt_{\rm em} = 0 \,.
\end{equation}
Rearranging this equation provides the fundamental relation between the coordinate time intervals at observation and emission
\begin{equation}
    \frac{dt_{\rm obs}}{dt_{\rm em}} = \frac{N(t_{\rm em})}{N(t_{\rm obs})} \frac{a(t_{\rm obs})}{a(t_{\rm em})} \,.
\end{equation}
Substituting the GCT scaling $N(t) = a^{b/4}$ and using $a(t_{\rm obs})=1$ (and $N(t_{\rm obs})=1$) and $a(t_{\rm em})=(1+z)^{-1}$, we obtain the modified time dilation law relating the coordinate times
\begin{equation}
    dt_{\rm obs} = a^{b/4} \cdot a^{-1} dt_{\rm em} = a^{-1+b/4} dt_{\rm em} = (1+z)^{1-b/4} dt_{\rm em} \,.
\end{equation}
Because the source is shielded, its intrinsic proper time $\tau_{\rm rest}$ coincides with the local coordinate interval at emission ($d\tau_{\rm rest} \approx dt_{\rm em}$ in the static gauge). Thus, the observed duration $\tau_{\rm obs}$ (measured by Earth observers where $N \approx 1$) relates to the intrinsic duration as $\tau_{\rm obs} \propto (1+z)^{1-b/4} \tau_{\rm rest}$.

This framework provides a unified interpretation for the disparate behaviors of astrophysical sources, which can be categorized into discrete transients and persistent variable sources. SNe Ia and GRBs represent the former. These events originate from deep within gravitationally bound systems—white dwarfs and collapsars, respectively. Due to the shielding mechanism, their intrinsic explosion physics is fixed ($\tau_{\rm rest} = \text{const}$). The observed time dilation is therefore a pure path effect, faithfully tracing the background metric scaling derived above.

In contrast, QSOs represent persistent sources where observations are subject to a critical selection effect. While the central engine of a QSO is also a gravitationally bound system, observations do not reveal a singular event with a fixed rest-frame duration. Instead, observations exhibit the stochastic variability of a continuous thermal accretion disk at a fixed bandpass. As will be derived in Section~\ref{sec:duration_scaling}, observing at a fixed wavelength selectively samples inner, faster dynamical regions of the disk at higher redshifts. This introduces an intrinsic time-shortening $\tau_{\rm intr} \propto (1+z)^{-2}$, which effectively cancels the geometric time dilation.

\begin{figure}[t]
\centering
\includegraphics[width=\linewidth]{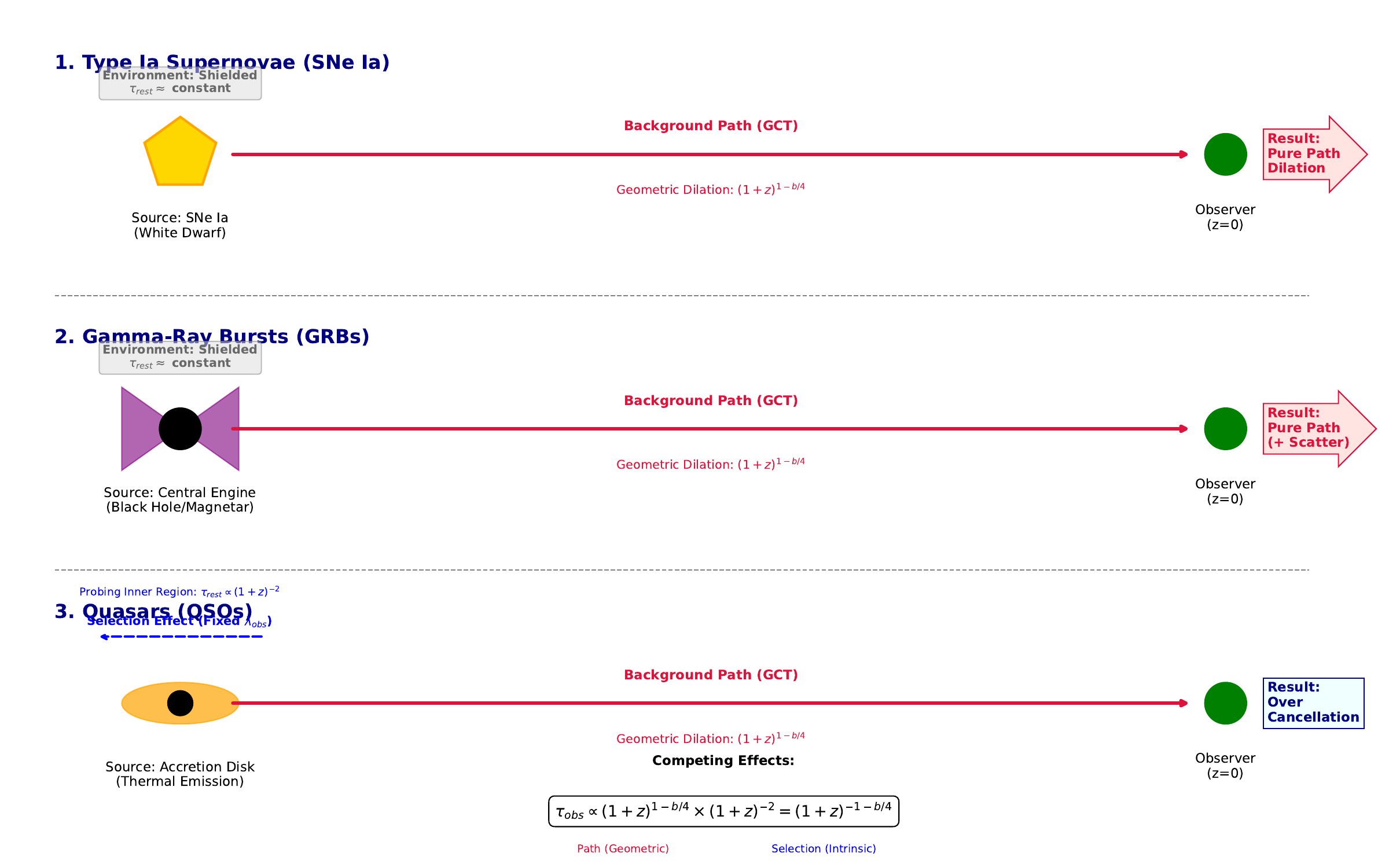}
\caption{Schematic representation of the unified GCT interpretation for three classes of cosmic clocks. \textbf{Top \& Middle:} SNe~Ia and GRBs originate from shielded, gravitationally bound progenitors. Their intrinsic timescales are fixed ($\tau_{\rm rest} = \text{const}$), so they act as standard clocks revealing the pure geometric path dilation $\tau_{\rm obs} \propto (1+z)^{1-b/4}$. \textbf{Bottom:} Quasars exhibit an apparent null result due to a selection effect. Fixed-wavelength observations probe inner, faster dynamical regions of the accretion disk at higher redshifts ($\tau_{\rm intr} \propto (1+z)^{-2}$), which effectively compensates the background time dilation signal.}
\label{fig:Triple_GCT}
\end{figure}

Figure~\ref{fig:Triple_GCT} schematically illustrates this unified interpretation. The top panel depicts SNe Ia, where the progenitor is shielded, and the light curve undergoes pure geometric stretching as it traverses the expanding medium. The middle panel shows GRBs, which operate under the same principle; despite their larger scatter, they trace the same path dilation. The bottom panel highlights the unique case of QSOs. Here, the redshifted observation window (blue to red shift) forces the observer to probe a hotter, inner region of the accretion disk (closer to the black hole) compared to a low-redshift counterpart. Because dynamical timescales are shorter at smaller radii ($\tau \propto R^{3/2}$), the intrinsic variability appears faster at high $z$, masking the cosmological signal.

\section{Cosmological Time Dilation in SNe, GRBs, and QSOs}
\label{sec:duration_scaling}

In the GCT framework established in Section~\ref{sec:theoretical_framework}, the observed duration of an astrophysical source is determined by the exact product of the cosmological background expansion and the intrinsic evolution of the source's internal clock. Unlike standard FLRW cosmology, where intrinsic timescales are assumed invariant by fiat, the present analysis explicitly separates the gauge-dependent background evolution from the local physics of gravitationally bound systems.

The relationship between the observed duration $\tau_{\rm obs}$ and the intrinsic rest-frame duration $\tau_{\rm rest}$ is rigorously defined as
\begin{equation}
  \tau_{\rm obs}(z) = \underbrace{(1+z)^{1-b/4}}_{\text{Path (Geometric)}} \times \underbrace{\tau_{\rm rest}(z)}_{\text{Source (Intrinsic)}} \,.
  \label{eq:master_scaling}
\end{equation}
The first term represents the accumulated time dilation due to photon propagation through the expanding background metric, governed by the generalized lapse function $N(t) \propto a^{b/4}$. As derived in Section~\ref{sec:theoretical_framework}, this path-dependent factor is universal for all extragalactic sources. The second term, $\tau_{\rm rest}(z)$, encapsulates the behavior of the local clock governing the source. Based on the principle of environmental shielding, $\tau_{\rm rest}(z)$ takes distinct forms depending on whether the source is a shielded bound system or subject to observational selection effects.

\subsection{Type Ia Supernovae: The Standard Shielded Clock}
\label{subsec:duration_SNe}

SNe Ia originate from the thermonuclear runaway of carbon-oxygen white dwarfs near the Chandrasekhar mass limit. The characteristic width of the supernova light curve is governed by the photon diffusion time through the expanding ejecta. Analytically, this timescale follows Arnett's rule~\cite{Arnett:1982ioj}
\begin{equation}
  \tau_{\rm diff} \approx \left( \frac{3 \kappa M_{\rm ej}}{4 \pi v_{\rm exp} c_{\rm local}} \right)^{1/2} \,,
  \label{eq:arnett}
\end{equation}
where $\kappa$ is the optical opacity, $M_{\rm ej}$ is the ejected mass, $v_{\rm exp}$ is the expansion velocity, and $c_{\rm local}$ is the local speed of light.

According to the environmental shielding principle, the white dwarf progenitor is a gravitationally bound system dynamically decoupled from the background GCT evolution. Consequently, the local physical parameters governing the explosion are fixed. The ejected mass remains near the Chandrasekhar limit ($M_{\rm ej} \approx 1.4 M_\odot$), the opacity $\kappa$ is determined by invariant atomic physics, and the local speed of light $c_{\rm local}$ remains constant at $c_0$ due to the static nature of the local metric ($b_{\rm local} \approx 0$). This implies that the intrinsic rest-frame duration is invariant across cosmic history
\begin{equation}
  \tau_{\rm SNe, rest}(z) = \tau_{\rm diff} = \text{const}.
\end{equation}
Substituting this into Eq.~\eqref{eq:master_scaling}, the observer-frame duration scales solely with the geometric dilation factor
\begin{equation}
  \tau_{\rm SNe, obs}(z) = \tau_{\rm diff} \, (1+z)^{1-b/4} \,.
  \label{eq:SNe_scaling}
\end{equation}

This prediction establishes SNe Ia as ideal background probes for the generalized geometry. While broad analyses of the Dark Energy Survey (DES) are broadly consistent with standard time dilation, high-precision constraints from specific observational bands provide a critical testing ground~\cite{Lee:2024kxa,Lee:2025vha}. Although the central values of certain bands slightly favor weaker dilation ($n < 1$), a generalized scaling with $b \approx -0.04$ (yielding an enhanced dilation exponent $n \approx 1.01$) remains well within the statistical error margins of the combined DES analysis. Such a negative value of $b$ is particularly compelling, as it is independently required to reduce the pre-recombination sound horizon and alleviate the Hubble tension~\cite{Lee:2025arX}. Thus, rather than simply confirming the standard model ($b=0$), high-precision SNe Ia data accommodate the non-trivial lapse function characterizing the GCT geometry, providing a permissible window for $b < 0$.

\subsection{Gamma-Ray Bursts: Shielded Engines with Astrophysical Scatter}
\label{subsec:duration_GRB}

Long GRBs are powered by the core collapse of massive stars, leading to the formation of a compact central engine—typically a black hole or a magnetar~\cite{Woosley:1993wj,MacFadyen:1998vz}. The fundamental timescale governing the engine activity is the dynamical timescale of the compact object. For a central engine of mass $M_{\rm engine}$ and characteristic radius $R$, the dynamical time is given by
\begin{equation}
  \tau_{\rm dyn} \approx 2\pi \sqrt{\frac{R^3}{G_{\rm local} M_{\rm engine}}} \approx \frac{2\pi R}{c_{\rm local}} \sqrt{\frac{R c_{\rm local}^2}{G_{\rm local} M_{\rm engine}}} \,.
  \label{eq:grb_engine}
\end{equation}
Since the central engine is a deep gravitationally bound system, the environmental shielding ensures that the local gravitational constant $G_{\rm local}$ and speed of light $c_{\rm local}$ are fixed at their laboratory values. Therefore, the intrinsic engine timescale is redshift-independent, $\tau_{\rm GRB, rest}(z) = \tau_{\rm dyn} = \text{const}$.

Following Eq.~\eqref{eq:master_scaling}, GRBs are predicted to obey the same scaling law as SNe Ia
\begin{equation}
  \tau_{\rm GRB, obs}(z) = \tau_{\rm dyn} \, (1+z)^{1-b/4} \,.
  \label{eq:GRB_scaling}
\end{equation}
However, unlike SNe Ia, the observed durations of GRBs (measured as $T_{90}$) exhibit significant scatter spanning orders of magnitude. This scatter arises from the diverse astrophysical properties of the progenitors—such as varying envelope masses, jet opening angles, and accretion rates~\cite{Piran:2004ba,Kumar:2014upa}—rather than from a deviation in the fundamental time dilation law. Thus, within the GCT framework, GRBs are classified as noisy shielded clocks, probing the same background geometry as SNe Ia but with lower precision due to intrinsic astrophysical variance.

\subsection{Quasars: Cancellation via Observational Selection}
\label{subsec:duration_QSO}

Quasar (QSO) variability studies have historically reported a null result, where characteristic timescales appear independent of redshift~\cite{Kelly:2009wy,MacLeod:2010qq,MacLeod:2011fn}. This behavior is resolved by recognizing that QSOs are continuous sources observed at a fixed observed wavelength $\lambda_{\rm obs}$, which introduces a redshift-dependent selection effect.

The emission from a quasar accretion disk is well-described by the Shakura-Sunyaev thin-disk model~\cite{Shakura:1972te,Frank:2002bo}. In this framework, the radial temperature profile $T(R)$ is determined by the release of gravitational potential energy as matter accretes onto a central supermassive black hole. For a black hole of mass $M_{\rm BH}$ and a mass accretion rate of gas $\dot{M}$, the temperature at a radial distance $R$ scales as
\begin{equation}
  T(R) \propto (M_{\rm BH} \dot{M})^{1/4} R^{-3/4} \,.
\end{equation}
Observation at a specific rest-frame wavelength $\lambda_{\rm em}$ probes the disk region where the local temperature dominates the thermal emission, a relationship approximated by Wien's law $T \sim \lambda_{\rm em}^{-1}$. Solving for the characteristic radius $R_\lambda$ corresponding to this effective emission temperature yields $R_\lambda \propto \lambda_{\rm em}^{4/3}$. This radius-wavelength relation is consistent with independent microlensing observations of accretion disk sizes~\cite{Morgan:2010xf}.

The intrinsic variability timescale is physically associated with the Keplerian dynamical time (or thermal timescale) governing the disk at this characteristic radius
\begin{equation}
  \tau_{\rm disk} \approx \sqrt{\frac{R_\lambda^3}{G M_{\rm BH}}} \,.
\end{equation}
Substituting $R_\lambda$ into this expression yields the dependence of the intrinsic duration on the emission wavelength
\begin{equation}
  \tau_{\rm disk} \propto R_\lambda^{3/2} \propto \left( \lambda_{\rm em}^{4/3} \right)^{3/2} = \lambda_{\rm em}^2 \,.
\end{equation}
In a fixed-bandpass survey, the rest-frame wavelength is related to the fixed observed wavelength by $\lambda_{\rm em} = \lambda_{\rm obs} / (1+z)$. Consequently, the intrinsic rest-frame duration probed at higher redshifts decreases as
\begin{equation}
  \tau_{\rm QSO, rest}(z) = \tau_{0} \left( \frac{1}{1+z} \right)^2 = \tau_{0} (1+z)^{-2} \,.
  \label{eq:QSO_intrinsic_exact}
\end{equation}

Finally, combining this intrinsic evolution with the geometric path dilation derived in Eq.~\eqref{eq:master_scaling}, the observed duration is given by
\begin{equation}
  \tau_{\rm QSO, obs}(z) = \tau_{0} (1+z)^{-2} \times (1+z)^{1-b/4} = \tau_{0} (1+z)^{-1-b/4} \,.
  \label{eq:QSO_final}
\end{equation}

For typical values of $b \approx -0.04$, the predicted exponent is approximately $-0.99$. It is acknowledged that this derivation assumes a leading-order scaling where parameters like accretion rate $\dot{M}$ and black hole mass $M_{\rm BH}$ do not evolve strongly with redshift in a way that perfectly cancels this effect. While astrophysical evolution may introduce scatter, the wavelength-dependent selection effect ($\tau \propto \lambda^2$) is a systematic geometric bias affecting all quasars. Remarkably, this prediction aligns with the widely reported null results in optical variability studies~\cite{Hawkins:2001be,Hawkins:2010xg}. A net scaling of $(1+z)^{-0.99}$ is consistent with a null result within observational scatter, whereas the standard $(1+z)^{+1}$ prediction is strongly excluded. Thus, the apparent lack of dilation in QSOs is naturally explained as an over-cancellation driven by the bandpass selection effect.  

This interpretation also resolves the apparent tension with the recent detection of time dilation in quasars reported by Lewis and Brewer~\cite{Lewis:2023jab}. By employing a Bayesian hierarchical model to standardize quasar light curves, they identified a signal consistent with $(1+z)$ expansion. However, their methodology relies on the assumption that the intrinsic variability characteristics, once standardized for luminosity and black hole mass, represent a redshift-independent clock. In the context of the GCT framework, their analysis effectively reconstructs the background geometric expansion (the first term in Eq.~\eqref{eq:master_scaling}) by statistically removing the intrinsic evolution. In contrast, historical studies reporting null results~\cite{Hawkins:2010xg} analyzed raw variability in fixed observing bands, where the bandpass selection effect ($\tau_{\rm intr} \propto (1+z)^{-2}$) is fully operative. Thus, the GCT framework accommodates both results: the selection effect masks the dilation in raw fixed-band observations, while the underlying geometric expansion—characterized by the shielded background metric—can be recovered if the intrinsic wavelength-dependence is explicitly corrected.

\subsection{Unified Picture: Summary of Duration-Redshift Scalings}
\label{subsec:unified_summary}

The results derived in the previous subsections demonstrate that the diverse observational signals of astrophysical sources can be understood as consequences of a single geometric expansion acting on different types of physical clocks. SNe Ia and GRBs behave as shielded clocks where $\tau_{\rm rest}$ is constant, revealing the pure background dilation. QSOs behave as bandpass-selected clocks where $\tau_{\rm rest}$ shrinks rapidly with redshift, masking the background dilation.

These theoretical scalings are juxtaposed in Table~\ref{tab:duration_scaling_summary}. The second column highlights the behavior of the intrinsic clock: it remains static for the shielded systems (SNe Ia and GRBs) but exhibits a strong inverse-square redshift dependence for the bandpass-selected system (QSOs). This intrinsic distinction drives the divergence in the observed scaling laws shown in the third column. While SNe Ia and GRBs act as faithful tracers of the background expansion geometry ($\tau_{\rm obs} \propto (1+z)^{1-b/4}$), quasars exhibit a net scaling where the selection effect dominates, reversing the expected time dilation trend.

\begin{table}[t]
\centering
\caption{Summary of observer frame duration scalings in the GCT framework.}
\label{tab:duration_scaling_summary}
\renewcommand{\arraystretch}{1.5}
\begin{tabular}{cccc}
\hline\hline
\textbf{Source Class}
& \textbf{Intrinsic Clock} $\tau_{\rm rest}(z)$
& \textbf{Observed Scaling} $\tau_{\rm obs}(z)$
& \textbf{Physical Origin} \\ \hline
SNe~Ia
& $\tau_{\rm diff} = \text{const}$
& $\propto (1+z)^{1-b/4}$
& Shielded Photon Diffusion \\

GRBs
& $\tau_{\rm dyn} = \text{const}$
& $\propto (1+z)^{1-b/4}$
& Shielded Engine (+ Scatter) \\

QSOs
& $\tau_{\rm disk} \propto (1+z)^{-2}$
& $\propto (1+z)^{-1-b/4}$
& Disk Dynamics + Selection Effect \\
\hline\hline
\end{tabular}
\end{table}

\begin{figure}[t]
  \centering
  \includegraphics[width=0.95\linewidth]{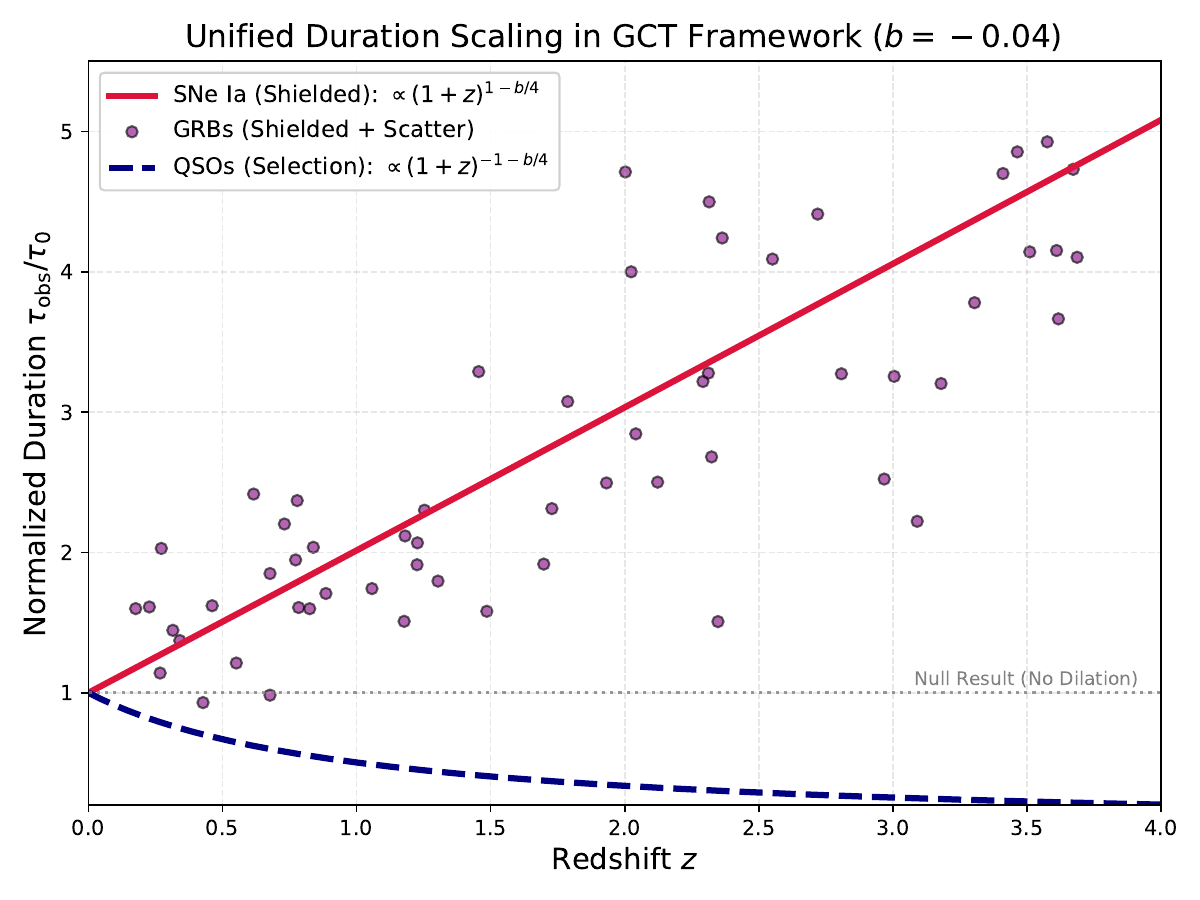}
  \caption{Conceptual comparison of the observed duration scaling with redshift for three source classes. The solid red line represents the pure cosmological time dilation ($ \propto (1+z)^{1-b/4}$) exhibited by shielded clocks like SNe Ia. The scattered dots represent a schematic illustration of GRBs, which follow the same trend but with significant astrophysical noise. The blue dashed line shows the predicted scaling for QSOs, where the intrinsic selection effect ($\propto (1+z)^{-2}$) over-compensates the dilation, resulting in a net decreasing trend ($\propto (1+z)^{-1-b/4}$). (Here $b=-0.04$ is assumed.)}
  \label{fig:duration_scaling_concept}
\end{figure}

Figure~\ref{fig:duration_scaling_concept} visually summarizes these competing effects and clarifies the unified interpretation. The solid red curve represents the baseline cosmological time dilation predicted by the GCT framework (with $b=-0.04$), which is closely tracked by SNe Ia. Since their intrinsic physics is shielded, they provide the cleanest measurement of the background metric expansion. The scattered data points representing GRBs follow the same mean relation, illustrating that while individual GRB engines possess significant astrophysical diversity, the population as a whole probes the same background path dilation as SNe Ia.

In sharp contrast, the dashed blue curve depicts the effective duration scaling for quasars. Due to the $\tau_{\rm intr} \propto (1+z)^{-2}$ selection effect derived in Sec.~\ref{subsec:duration_QSO}, the net observed duration scales approximately as $(1+z)^{-1}$. This results in a decreasing trend with redshift, fundamentally distinct from the increasing trends of SNe and GRBs. This graphical comparison demonstrates that the null or negative results often reported in quasar variability studies are not evidence against cosmic expansion, but are fully consistent with the over-compensation predicted by the thermal accretion disk physics under fixed-wavelength observations.

\section{Discussion}
\label{sec:discussion}

In this work, I demonstrate that the apparent conflict between SNe~Ia, GRB, and QSO time dilation signals can be resolved within the GCT framework. A central feature of this analysis is that the parameter $b$ is treated as an externally constrained parameter, consistent with values ($b \approx -0.04$) derived from independent cosmological tension studies~\cite{Lee:2025arX,Lee:2025vha}. Rather than fitting $b$ freely to transient data, I show that this single geometric parameter, motivated by the $H_0$ tension, naturally reproduces the diverse time-domain behaviors of astrophysical sources without additional fine-tuning. The results presented here should be interpreted in the broader context of recent studies that have progressively clarified the geometric origin of CTD, understood as an observational manifestation of the gauge-dependent normalization of cosmological time.  For clarity, the role of the parameter $b$ should be distinguished from models that alter the background expansion dynamics. Throughout this work, $b$ enters only through the temporal gauge choice and affects observable time dilation exclusively through photon propagation, while the Friedmann equations and matter evolution remain unchanged.

Accordingly, the apparent impact of the GCT framework is confined to the
operational definition of cosmological time, not to the background dynamics
themselves. Standard distance observables, including the comoving distance and
sound-horizon-related quantities, depend on the invariant combination $c/H$ and
therefore retain exactly the same functional form as in the standard FLRW cosmological model.
In this sense, the GCT framework reorganizes how cosmological time is measured and
interpreted, without altering the underlying expansion history.

Previous analyses have demonstrated that the mild redshift dependence reported in the locally inferred Hubble parameter from SNe~Ia data can be consistently interpreted as a consequence of this gauge freedom, rather than as evidence for modified expansion dynamics. Within this interpretation, departures from the standard $(1+z)$ time dilation relation arise naturally from the same lapse function freedom ($N(t) \propto a^{b/4}$) that underlies the GCT framework. From this perspective, the compatibility between the GCT prediction $\tau_{\rm obs} \propto (1+z)^{1-b/4}$ and existing measurements of supernova time dilation implies that SNe~Ia act as faithful tracers of the background geometry. Recent analyses of SNe~Ia samples from the DES have demonstrated overall consistency with the standard $(1+z)$ scaling, while the statistical uncertainties fully accommodate the slight enhancement ($n \approx 1.01$) predicted by $b \approx -0.04$. This highlights the internal consistency of the framework: a single parameter $b$ motivated by early-universe physics (the $H_0$ tension) is observationally viable in late-universe time-domain measurements, reproducing standard observations without invoking new dynamical degrees of freedom.

An important implication of this work concerns the physical interpretation of GRBs. Historically, the weaker and scattered evidence for time dilation in GRBs has been attributed to complex engine evolution or observational biases. However, under the environmental shielding principle proposed here, the central engines of GRBs—being deep gravitational potential wells—are just as shielded as SNe~Ia progenitors. This suggests that the fundamental scaling of GRBs is indeed identical to that of SNe~Ia ($\propto (1+z)^{1-b/4}$). The observed scatter in $T_{90}$ durations is therefore interpreted not as a failure of time dilation, but as astrophysical noise superimposed on a clean geometric signal. This reframes GRBs from being outliers to being noisy standard clocks that support the same metric expansion law as SNe~Ia.

The most distinct resolution offered by this framework addresses the long-standing puzzle of quasar variability. The apparent conflict between the null results of structure-function analyses (e.g., Hawkins; OGLE) and the recent detection of time dilation by Lewis and Brewer~\cite{Lewis:2023jab} is resolved by examining how the intrinsic clock is defined and observed. The analysis of Lewis and Brewer explicitly reconstructs rest-frame variability timescales through hierarchical modeling. By mathematically removing the bandpass dependence and standardizing the light curves, they effectively probe the shielded physics of the source, thereby recovering the underlying geometric time dilation signal consistent with expansion. This result is fully consistent with the GCT prediction for a path-dependent observable when intrinsic evolution is accounted for.

In contrast, traditional analyses such as those by Hawkins measure variability at a fixed observed wavelength. As derived in Section~\ref{sec:duration_scaling}, this introduces a powerful selection effect where $\tau_{\rm intr} \propto (1+z)^{-2}$. This intrinsic shortening competes with and effectively cancels the geometric cosmological time dilation. Therefore, the absence of a signal in fixed-bandpass studies does not contradict the expansion of the universe; rather, it confirms the physical model of thermally stratified accretion disks interacting with the redshifted observation window. The GCT framework thus unifies these seemingly contradictory results: Lewis and Brewer detect the pure path effect by correcting for source physics, while Hawkins detects the net result of the path dilation masked by the selection effect. Both are correct within their respective observational definitions, and both are predicted by the unified equations derived in this work.

Taken together, the diverse behaviors of SNe, GRBs, and QSOs reinforce a single coherent picture. The universe expands according to a generalized metric ($b \approx -0.04$). Local bound systems, including SNe progenitors and GRB engines, are shielded against this expansion ($b_{\rm local} \approx 0$). When observed directly as shielded clocks (SNe, GRBs), the background dilation is revealed. When observed through the filter of a fixed bandpass (QSOs), the signal is masked by intrinsic selection effects. This unified interpretation elevates transient duration measurements from a mere consistency check to a sophisticated probe of how cosmological geometry interacts with local astrophysical environments.

\subsection{Relation to Look-back Time Approaches to the $H_0$ Tension}

A number of recent studies have emphasized the role of cosmological look-back time
in alleviating the $H_0$ tension by reinterpreting how cosmic ages are inferred at
different redshifts~\cite{Capozziello:2023ewq,Vagnozzi:2023nrq,Capozziello:2024stm,Hu:2025fsz}.
In these approaches, the central observation is that the inferred Hubble constant
depends sensitively on how observational time intervals are mapped onto the cosmic
expansion history, particularly when integrated age constraints or cosmic
chronometers are involved.

The framework developed in this work is complementary to these look-back time
approaches rather than alternative. While look-back time analyses focus on the
interpretation of integrated cosmic ages accumulated along the expansion history,
the present work addresses a more elementary and operational question: which
astrophysical processes constitute physical clocks that directly trace the
background geometry. In this sense, the cosmological look-back time can be viewed
as an integrated consequence of the same generalized temporal structure discussed
here at the level of local time-dilation observables.

It is important to emphasize that this complementarity does not arise from any
modification of the background expansion. In the Generalized Cosmological Time
framework, the integral structure of distance and conformal time remains unchanged
and identical to that of the standard FLRW cosmological model. Rather, the distinction
arises from how different physical clocks sample the same underlying geometry when
mapped onto observational time variables. From this perspective, generalized
interpretations of look-back time can be understood as phenomenological
manifestations of the same temporal gauge freedom explored here.

Specifically, the Generalized Cosmological Time framework clarifies that different
astrophysical probes may couple differently to the background time slicing,
depending on whether their intrinsic clocks are shielded from or exposed to
environmental and observational selection effects. This distinction precedes and
underlies any interpretation based on integrated look-back time and provides a
microscopic explanation for why different probes can yield apparently inconsistent
inferences of the cosmic expansion rate, even when they probe the same underlying
geometry.

\section{Conclusion}
\label{sec:conclusion}

This work presents a unified physical interpretation of cosmological time dilation signals observed across Type~Ia supernovae, Gamma-Ray Bursts, and Quasars within the framework of Generalized Cosmological Time. By identifying the geometric origin of the generalized lapse function, I demonstrate that the disparate behaviors of these sources are not contradictions but consistent manifestations of how different physical clocks interact with the expanding background metric.

A central outcome of this analysis is the classification of transients based on the principle of environmental shielding. I argue that gravitationally bound systems—specifically the white dwarf progenitors of SNe~Ia and the compact central engines of GRBs—are dynamically decoupled from the global Hubble flow. Within these deep potential wells, local physical constants are shielded against the background cosmological time evolution. Consequently, both SNe~Ia and GRBs act as standard, or near-standard, clocks where the intrinsic duration remains static ($\tau_{\rm rest} \approx \text{const}$). The observed time dilation for these sources arises purely from the accumulated photon flight time through the expanding background metric. This predicts a universal scaling $\tau_{\rm obs} \propto (1+z)^{1-b/4}$, which is fully consistent with precision measurements of SNe~Ia from the Dark Energy Survey. Furthermore, this framework reinterprets GRBs not as anomalous outliers with evolving engines, but as noisy standard clocks that follow the same fundamental geometric scaling as SNe~Ia, albeit obscured by significant astrophysical scatter.

In sharp contrast, the long-standing puzzle of the null result in quasar variability is resolved by accounting for the observational selection effect inherent to fixed-wavelength surveys. This analysis demonstrates that for thermal accretion disks, observing at a fixed optical bandpass inevitably probes progressively inner, faster-rotating regions at higher redshifts. The resulting intrinsic time-shortening effect ($\tau_{\rm intr} \propto (1+z)^{-2}$) is sufficiently strong to over-compensate the background cosmological time dilation. This mechanism naturally predicts a net observed duration that is flat or even decreasing with redshift, explaining why traditional structure-function analyses yield null results without implying a breakdown of relativistic expansion. The unified picture is further supported by the recent detection of time dilation in quasars when rest-frame timescales are explicitly reconstructed, confirming that the background geometric dilation is present but masked by selection effects in standard observations.

Ultimately, this framework elevates cosmological time dilation from a qualitative consistency check to a precision probe of the metric structure of spacetime. The GCT parameter $b$ quantifies the normalization of the cosmological time gauge, offering a new dimension to test general relativity on cosmological scales. Future time-domain surveys, by simultaneously measuring the clean geometric signal from SNe~Ia and disentangling the complex selection effects in QSOs, will be critical in constraining this generalized geometric freedom and determining the true nature of cosmic time.

\begin{acknowledgments}
This work is supported by the Basic Science Research Program through the National Research Foundation of Korea (NRF), funded by the Ministry of Science and ICT under Grant No.~NRF-RS-2021-NR059413 and NRF-2022R1A2C1005050.
\end{acknowledgments}



\end{document}